**PAPER • OPEN ACCESS**

# The application of geographic information systems in schools around the world: a retrospective analysis

To cite this article: I Kholoshyn et al 2021 J. Phys.: Conf. Ser. **1840** 012017

View the article online for updates and enhancements.

# The application of geographic information systems in schools around the world: a retrospective analysis

**I Kholoshyn[1], T Nazarenko[2], O Bondarenko[1], O Hanchuk[1] and I Varfolomyeyeva[1]**

[1] Kryvyi Rih State Pedagogical University, 54 Gagarin Ave., Kryvyi Rih, 50086, Ukraine

[2] Institute of Pedagogy of the National Academy of Educational Sciences of Ukraine, 52-D Sichovykh Striltsiv Str., 04053, Kiev, Ukraine

E-mail: bondarenko.olga@kdpu.edu.ua

**Abstract.** The article is devoted to the problem of incorporation geographic information systems (GIS) in world school practice. The authors single out the stages of GIS application in school geographical education based on the retrospective analysis of the scientific literature. The first stage (late 70 s – early 90s of the 20[th] century) is the beginning of the first educational GIS programs and partnership agreements between schools and universities. The second stage (mid-90s of the 20[th] century – the beginning of the 21[st] century) comprises the distribution of GIS-educational programs in European and Australian schools with the involvement of leading developers of GIS-packages (ESRI, Intergraph, MapInfo Corp., etc.). The third stage (2005–2012) marks the spread of the GIS school education in Eastern Europe, Asia, Africa and Latin America; on the fourth stage (from 2012 to the present) geographic information systems emerge in school curricula in most countries. The characteristics of the GIS-technologies development stages are given considering the GIS didactic possibilities for the study of school geography, as well as highlighting their advantages and disadvantages.

## 1. Introduction

*1.1. Scientific relevance of the research*
The humanity has made a gigantic step from industrial to information era by the end of the second millennium. Nowadays the decisive factors in the development of society are intelligence and information access compared to material and labor resources of the previous epochs. Presently the information of spatiotemporal nature is mainly used for solution of practical tasks [26], [27], [34]. Accordingly, an objective necessity arises to bring in geographic information systems that provide full-fledged and operative analysis of spatiotemporal data [22], [25], [33].

　　Geographic information systems (GIS) in a broad sense mean a model of the real world, and in a narrow sense – modern computer technology for mapping and analysis of natural and anthropogenic objects, phenomena, processes and events occurring on our planet [38].

　　William J. Cook, Warren Cohen et al considers GIS to be one of 25 most important events that caused the most substantial impact on people's life in 20[th] century [12]. In modern geographic science, GIS-technologies play exactly the role that in the era of the Great Geographical Discoveries vehicles played [7].

Today GIS become not only an integral part of professional work place in various fields of science and economy but surround people in day-to-day life. Thus, the younger generation urgently need to master this type of geographic information technology. Besides, the GIS introduction in school education allows to create an exciting and potentially favorable learning environment. Therefore, we believe that GIS education should take its worthy place as a method of world exploration.

*1.2. Article objective*
The effective use of GIS in school geographical education is impossible without knowing the history of their development, the use of advanced pedagogical experience and a balanced analysis of the problems that happened and still occur during their implementation. The purpose of the proposed publication is to give a retrospective analysis of the use of GIS in global school practice.

## 2. Results and discussion
We consider it primarily important to identify the main stages of GIS technologies development in school geographical education for the achievement of the article goal.

*The first stage* of GIS-technologies development in school geographical education occurs in the late 70's – early 90's of the twentieth century, and is associated with the emergence of the first training GIS programs in the United States and Canada. The introduction of GIS in geographical education dates back to the end of 1970s at the university level and since then the interest in them grows like an avalanche. Thus, the number of geographic GIS curricula offered by American and Canadian universities in 1984, according to David Richard Green, was approximately 10 [20]. By the end of 1990, their number exceeded 2,000, and the disciplines incorporating GIS included History, Information Technology, Biology, Mathematics and other sciences.

Such active use of GIS in universities ultimately provided a powerful impact on the secondary education system. The interest in this field at the school level grew dramatically after numerous studies proving that GIS is an efficient educational instrument. They are made by Ronald F. Abler [1], Richard H. Audet and Joshua Paris [2], Thomas R. Baker and Sarah Witham Bednarz [3], Nick Bearman, Nick Jones, Isabel André, Herculano Alberto Cachinho and Michael DeMers [5], Sarah Witham Bednarz and Joop van der Schee [6], Josep Blat, Angel V. Delgado, Maurice Ruiz and Joana Maria Seguí [9], James Cadoux-Hudson and Ian Heywood [11], William J. Cook, Warren Cohen et al [12], David DiBiase [16], Karen K. Kemp, Michael F. Goodchild and Rustin F. Dodson [23], Joseph J. Kerski [24], Carol McGuinness [32], Jonathan Raper and Nick Green [35], Jan Ketil Rød, Wenche Larsen and Einar Nilsen [36], Tim Sutton, Otto Dassau and Marcelle Sutton [39], Robert F. Tinker [40], Patrick Wiegand [42]. Accordingly, GIS merged in school curricula in many countries, among which we note the United States and Canada, and among European countries – Great Britain, Denmark, Germany, France, Finland, Sweden and the Netherlands. The United States and Canada take the lead in this area as they were the first to use GIS in high school in the 1990s. Virtually, they laid the foundation for the development of school geo-informatics not only in the United States, but also stimulated its development around the world.

National Geographic Society of America (NGSA) was one of the first to be involved in the work of introducing GIS in school education process. NGSA began to hold introductory seminars with school teachers throughout the country in 1986, where teachers received basic skills in working with new educational resources.

The National Center for Geographic Information Analysis (NCGIA) carried out the planned and purposeful development of GIS education in US schools. Established in 1989, the Center is a consortium of three universities (University of California, Santa-Barbara; Maine State University and Buffalo University of New York State), which was funded mainly by the National Science Foundation. Its mission was to conduct outreach activities aimed at the development of GIS technologies, among both professionals and a wide range of the persons concerned. The early stages of this work were reflected in the Yearbook of the Association of Geographical Information 1992–1993 [11].

Secondary Education Project (SEP) developed by NCGIA, aimed at solving the following problems:

- to favor dialogue between a teacher and students;
- to collect, develop and analyze GIS teaching materials;
- to inform of current and future GIS schools' activities;
- to deliver information;
- to encourage development of the relevant GIS teaching materials for schools.

One of the first activities was the development of a weekly GIS seminar for secondary school teachers, which later became a series of "GIS in schools" seminars aimed at:

- introduction to GIS;
- studying work GIS applications;
- practical work with GIS software;
- discussion the role of GIS at schools;
- analysis of available teaching materials.

The researchers from the Institute for Environmental Systems Research (ESRI), established by Jack Dangermond in 1969, facilitated a special development of geo-information school education. ESRI established a school and library department in 1992 in order to gradually use GIS in schools in the United States and the United Kingdom. The mission of the department is to form a GIS-supported spatially educated society. Since 1993, ESRI has been publishing the ArcSchool Reader newsletter. In addition to geo-informational news, this information bulletin supplies the instructions for using GIS ArcView, one of the main products of the institute, in various disciplines of the educational process: from pre-school to high school. Generally, the educational activity of ESRI staff in almost all countries have contributed to the involvement of a wide range of educators in the GIS community. The ArcView program has become the most widespread in the field of education.

Due to the widespread of the Internet, ESRI initiated interactive education of teachers and students at various levels and majors related to the GIS-education.

Robert F. Tinker demonstrated the practical application of GIS possibilities for secondary education system in his article in 1992 [40]. The author described personal experience of using digital maps in the study of various environmental issues by four-six grade students in the framework of the KidNet project. Tinker demonstrated the educational importance of GIS application to the analysis of attributive data of the environment monitoring, obtained during surface and satellite observations. The students involved in the project saw in a live and spontaneous way what opportunities are revealed in the spatial analysis with the involvement of GIS. The author paid special attention to the motivation factor that stimulated the students' interest for acquiring new, previously unknown and unused technologies.

The emergence of new scientific directions is inevitably accompanied by a surge of professional meetings, conferences and congresses on the subject. Geographic information systems in education are no exception. Thus, in January 1994, the US National Geographic Society hosted the first annual conference on the educational potential of GIS. The conference, announced as EdGIS, was a great success and became an annual event. EdGIS still makes a significant contribution to solving the problems of GIS implementation in school education: methodology, curriculum, psychology, etc.

The effectiveness of EdGIS contributed to the emergence of new conferences at various levels: the first international conference on GIS education (GISED'98), GIS education: the European perspective (EUGISES'98) and a number of others.

The IDRISI geographic information package from Clark Labs Corporation (USA) is considered successful. Since 1995, more than 250 copies of the program have found their way into primary and secondary schools around the world. To provide informational support for the implementation and use of their product, Clark Labs staff regularly conducts workshops for teachers in the United States, Canada, and other countries.

GIS education, as mentioned above, originated in universities. For example, back in the 1970s, the pioneers in this field were the Computer Graphics Laboratory at Harvard University (USA) and the Department of Experimental Cartography at the Royal College of Art (UK). In these institutions new computer algorithms and programs designed for spatial information processing were developed experimentally by trial and error method. Due to this, advanced students and teachers gained the first experience of integrating GIS technologies into the educational process, which began to spread actively among other universities in the world.

Thus, in 1995, a joint project of hundreds of American universities "Environmental and Spatial Technologies" (EAST) was launched. The project uses strategies and technologies of problem-based learning to stimulate the intellectual development of students based on GIS.

Since then, GIS technologies began to evolve so rapidly that educational opportunities simply have not kept pace with them. GIS software products began to provide profits and buy out quickly. Many educational courses have been developed, the content of which does not involve the study of basic concepts of GIS, but were mainly aimed at studying the use of certain GIS. In the mid-1990s, it became clear that there was a shortage of specialists capable of conveying basic geo-information knowledge to future teachers. This contributed to the development of the NCGIA GIS Training Course by the US National Science Foundation. This project was based on the premise that the developed teaching materials would be widely disseminated among GIS teachers. The core of the course, about 1000 pages, was translated into many languages and purchased by many educational institutions around the world (more than 70 countries purchased about 1300 copies of the analyzed training course in 5 years). According to an international survey, more than 450 universities in the United States, Europe, and Australia provided GIS education by the 1990s [28].

Consequently, partnership agreements between schools and higher education institutions made a significant contribution to solving the problems of primary GIS education. For example, the GIS Laboratory of Washington College developed a curriculum for studying geographic information systems by students of all ages. This program allowed students to gain basic GIS skills. The recommended form of the educational process is distance learning (online) and weekly courses in summer camps.

Researchers of the University of Texas, in collaboration with the National Science Foundation (NSF) and Covington High School (Texas, USA), developed a curriculum including the comprehensive use of GIS in the Geography for students of six-eight grades. The authors insist that the age level and educational process stages in high school allow to combine optimally the standard curriculum with geospatial technologies. Thus, in the sixth grade school curriculum, students acquire new environmental knowledge. The use of GIS at this level helps collect and organize information flows effectively. At this stage, students develop basic skills and abilities to work with GIS software (figure 1).

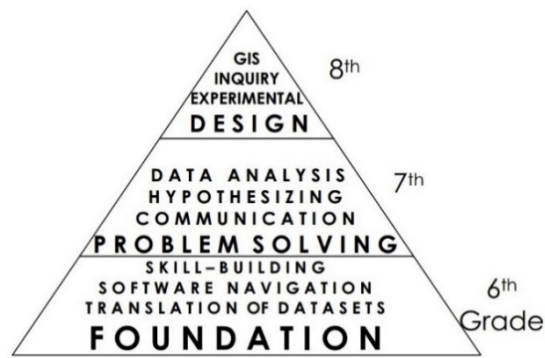

**Figure 1.** Structural diagram of GIS teaching program at Covington Secondary School, Texas, USA [31].

In the seventh grade the role of GIS analytical functions increases in view of the growing students' self-awareness. Students learn to perform basic spatial analysis at this level, which ultimately contributes to their ability to produce hypotheses to solve practical problems.

In the eighth grade, students are ready to solve intellectual problems by doing research projects and experiments. They study to use GIS for construction predictive models that have significant practical value.

*The second stage* of GIS technologies development in geographical school education falls on the middle of the 1990s – the beginning of the 21st century. This stage is characterized by the spread of GIS curriculums in schools of the most developed countries in Europe and Australia.

In Europe, geographic information educational systems were first used in schools in the United Kingdom. For example, teachers at Bishop's Stortford College under the leadership of Peter O'Connor developed a multi-level program for the use of GIS in schools as part of the Geographical Association project in the early 1990s. The program is based on the GIS package of ESRI ArcView (version 9.1) and covers three conceptual levels around which GIS-educational activities were organized:

1 level. Spatial data presentation.
2 level. Spatial data processing and analyzing.
3 level. Spatial data input and editing.

At the first level, GIS is used as a tool that allows teachers to demonstrate GIS capabilities in the processing of various spatial information to students aged from 9 to 13 (see table 1).

**Table 1.** Examples of activity types and levels of student academic achievements at the first level of GIS study.

| Level of GIS theory knowledge | GIS software knowledge requirements | Studying examples |
|---|---|---|
| None | Skills to start GIS app; Skills to use basic GIS-pack navigation functions (for instance, scaling and panning); Skills to work with data layers | 9 years: Demographic indicators; quality of life measures; indicators of regional development; crime development models in the UK. 10 years: Use of global energy; energy resources in the UK, 11 years: Global warming; sea level; distribution of earthquakes and volcanoes on planet Earth. 12 years: Social and economic structures of the city of Cambridge. 13 years: Industries nesting; patterns of world derivation distribution |

This approach does not require students to know the theory of GIS unless they have some basic software skills. For example, a teacher demonstrates for 9-year-olds the layers of cartographic studies of crime patterns in England and Wales (figure 2). So, students get an idea of the different nature of crime in rural areas compared to urban ones.

The nature of the processed spatial information becomes more complicated with the age of students. The purpose of the first stage is considered fulfilled if students have learned to obtain information that is not available when analyzing statistical tables and graphs.

At the second level, students need a deeper understanding of GIS theory, the principles of cartographic and statistical analysis (see table 2).

Teachers plan and organize work in such a way that provides a gradual increase in students GIS knowledge. It is important that at this level students already acquire a certain level of GIS theory. The higher this level is; the more complex analytical tasks they are able to solve.

Consequently, students gain greater academic independence and can take full advantage of the program's wide range of analytical capabilities. GIS at the second level allows students not only to determine where certain processes (phenomena) occur, but, more importantly, to answer why they take place there. Figure 3 shows an example of a cartographic analysis of the distribution of infant mortality in Europe and Africa, performed using GIS by 13-year-old students. They identified not only patterns in the spatial disproportion of this indicator, but also established the factors that determine it.

Teaching students to input and edit spatial data into GIS at the third stage is, perhaps, one of the most important skill formed within the program framework (see table 3).

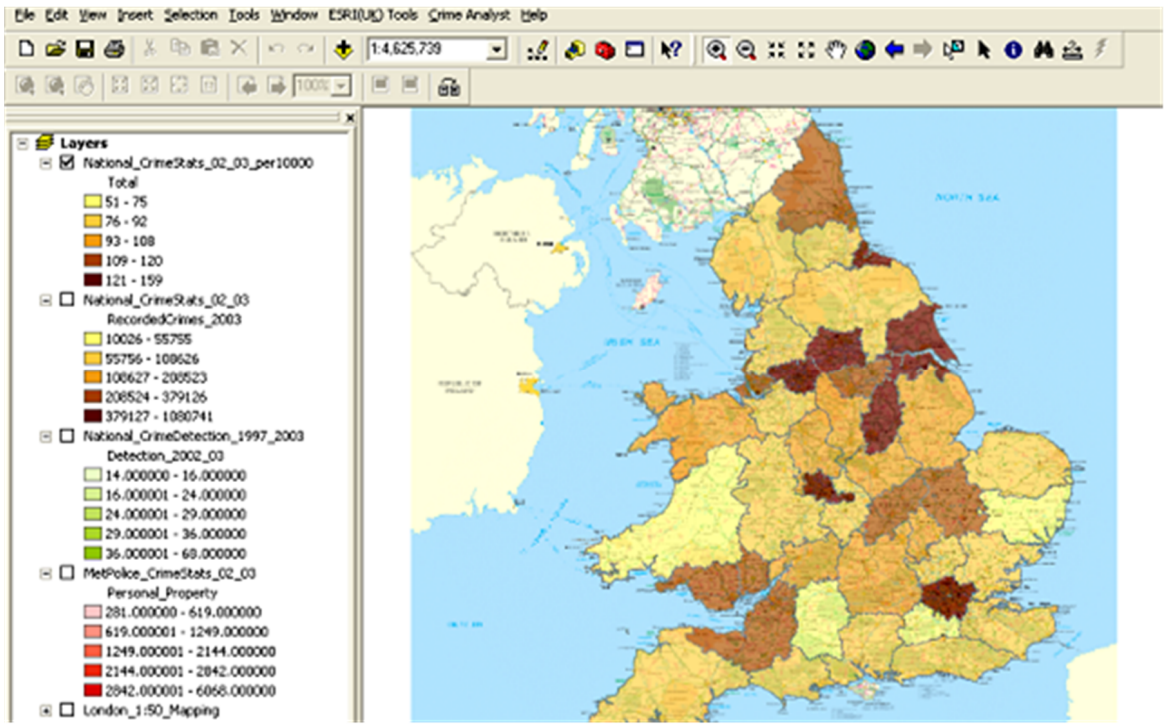

**Figure 2.** The example of the map demonstrating the patterns of crime distribution in England and Wales [8].

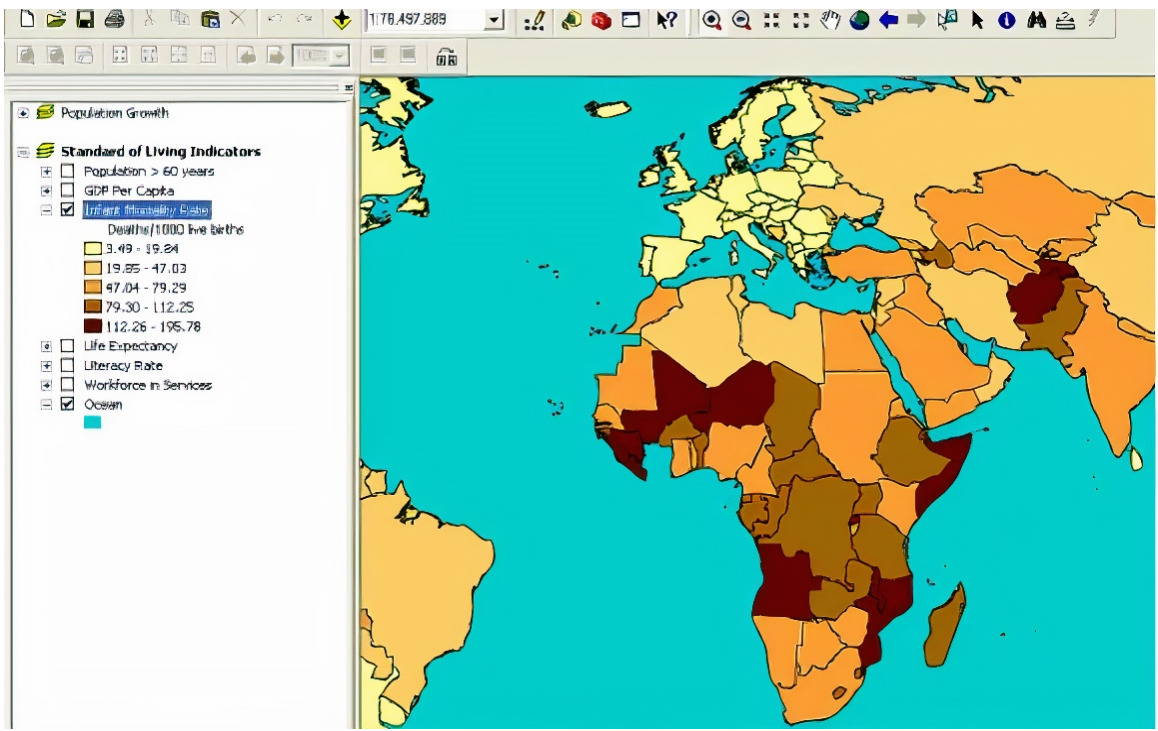

**Figure 3.** The example of the map demonstrating infant mortality indicator distribution pattern in Europe and Africa [8].

Table 2. Examples of activity types and levels of student academic achievements at the second level of GIS study.

| Level of processing and data analysis skills | Level of GIS theory knowledge | GIS software knowledge requirements | Studying examples |
|---|---|---|---|
| Basic | Qualitative and quantitative methods of maps classification; map symbolization and using of a size, a form, color shade, color and texture brightness. Two-dimensional data display methods. Map planning and designing. Object length, perimeter and square measuring. | Students are to develop understanding of classification range, symbolization and display designing methods available in the software and how to access them. Students should practice using GIS input data in forms of maps, images, graphs, tables and reports. | 9 years: Spatial representation of historic and future of Bishop's Stortford development. 12 years: Spatial representation and global tectonic activity symbolization |
| Intermediate | Data and queries selection. Base logic principles. Data aggregation methods. Statistics methods. 3D-cartography and visualization making. | Students should be familiar with statistics and data processing capabilities in GIS packages | 10 years: Identifying areas of high and low levels of economic development on a global scale. 13 years: Mapping of seismic activity zones in the USA |
| Advanced | Reclassification. Buffering and zoning. Overlay operations. Spatial interpolation. Surface analysis. | | 13 years: GIS analytic functions are most often used together with experienced users within GIS-projects of A1 level. |

These skills are specially significance if students plan to apply GIS in further education or profession. Practically, at this stage student learn various methods of cartographic and attributive data input, their vectorization and identification.

Table 3. Examples of activity types and levels of students' learning achievements at the third level of GIS study.

| General knowledge and GIS theory level | Teaching examples |
|---|---|
| Data input into electronic tables. Table data Import in GIS-pack. Making dotted, linear and areal cartographic layers. Editing dotted, linear and areal cartographic layers. Bonding attributive data with dotted, linear and polygonal cartographic layers. Understanding the role of cartographic projections and coordinates systems in GIS. | 11 years. GIS-project: representation of Bishop's Stortford municipal structure 13 years. Students' work in GIS-projects of A-2 level. |

Figure 4 shows an example of the map made by students within A2 project, aimed to make a forecast the results of the 2004 elections within Haolow County, as a representation of spatial disproportion in the level of population deprivation. While working, students went through all stages of GIS, starting with the collection and input of primary information and ending with the compilation and analysis of cartographic materials, thus demonstrating the skills formed at all three conceptual levels of the program.

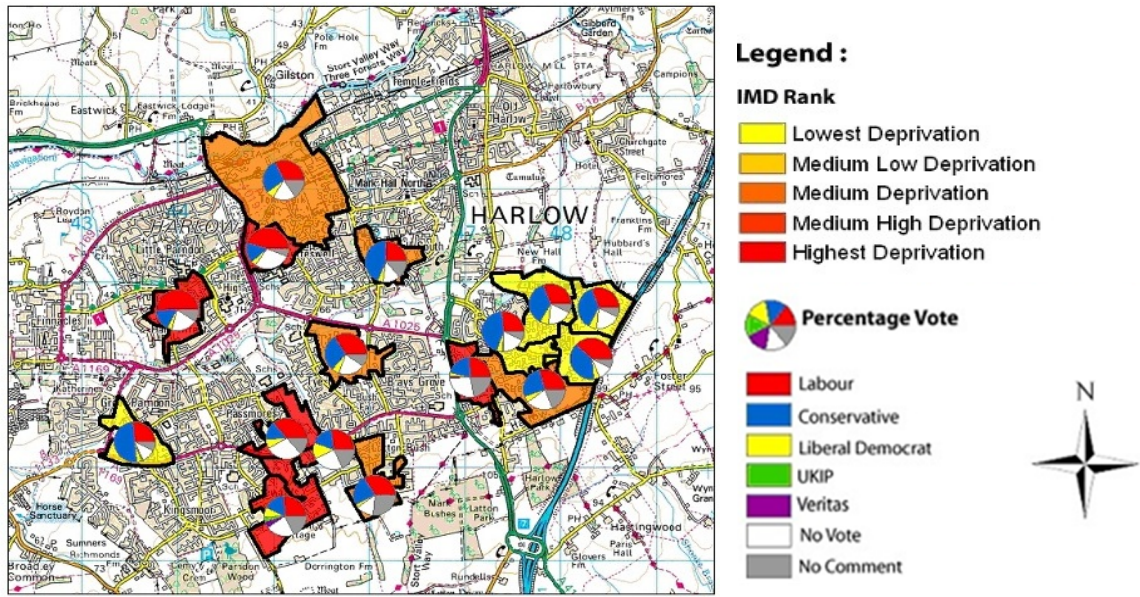

**Figure 4.** The map of 2004 election results forecast within the Harlow County, USA (Green, 2000).

In 1999, ESRI Schools and Libraries, together with highly qualified teachers and methodologists, developed the High School GIS Study Program, which is now widely used in thousands of schools around the world. The program is based on the use of GIS ArcView. In order to expand the GIS community, ESRI is creating its own virtual campus to study GIS technology [19]. The campus currently has 200,000 registered users from 185 countries.

That same year, ESRI, together with the World Resources Institute (IWR), published an ArcView GIS extension called DataScape, which allowed students to use an IWR database covering more than 450 socioeconomic indicators from more than 160 countries.

GIS became especially popular among students who performed thematic tasks of special scientific and practical significance. An example is the study of students in Minnesota (USA) who gained important information about the migration of endangered wolf species via GPS collars and GIS used. In Chelsea (Massachusetts, USA), students used GIS to identify storage areas for environmentally hazardous materials in the region. These materials were used by the State Emergency Management Agency [1]. In a summer camp research program in Ohio, students used GIS to study watersheds and to model the development of ecological processes. Young researchers also received information about the ecology of streams and soil erosion as part of their work.

Despite extensive research conducted by individual educational and commercial organizations from 1990 to 2000, we note that at the beginning of the 21$^{st}$ century GIS were disregarded in the school education system globally. For example, in the United States, the country with the most active development of geographic information technology, less than 50% of schools used GIS in 2003. When we assess the level of GIS use in European secondary schools at the end of the 20$^{th}$ century, we found out that it was about the same in the UK as in the US and Canada, and did not exceed 20% in France, Sweden and Finland. In other European countries, these figures are even lower: in Russia, Ukraine, Turkey and most others, GIS was absent in school curricula at all and was used only in individual educational institutions thanks to enthusiastic teachers.

Active and comprehensive implementation of GIS in the world schools began at the beginning of the 21$^{st}$ century. The analyses of the current use of GIS in world schools reviles that the historical leadership of the United States in the development of GIS education mirrors in the use of GIS in secondary schools.

Undoubtedly, the main factor is the ESRI policy of GIS education priority in schools, colleges and universities. For example, in the United States, the ESRI Yearbook for Teachers and a special GIS handbook are published, and training seminars are held. Websites are created to store information related to the implementation of GIS in the curricula of different countries: the ArgGIS Online for Schools site [39] offering software, digital maps, lesson plans, ESRI products instructions, etc. According to education surveys, 47% of GIS study groups use ESRI products. Today, the vast amount of U.S. geography teachers uses spatial information analysis programs as an established educational tool.

Other countries, even the most developed ones such as France, Germany and the United Kingdom, lag far behind in this regard. Thus, only since 2007, according to educational qualification programs, GIS has become mandatory for high school students in the UK. The Geography curriculum requires students to master spatial thinking skills, use maps, electronic visual images, the modern technologies, including GIS, to obtain and analyze information.

The German education system comprises several programs of Federal Lands with different educational content and terms of training. Despite attempts to integrate these programs into a single system, there are still some differences. Still GIS is introduced in most schools in Germany, although at different times and with different approaches. Diercke GIS software based on ArcView 3.2 was developed especially for educational purposes. The package comes with a large collection of maps and ready-to-use data (figure 5).

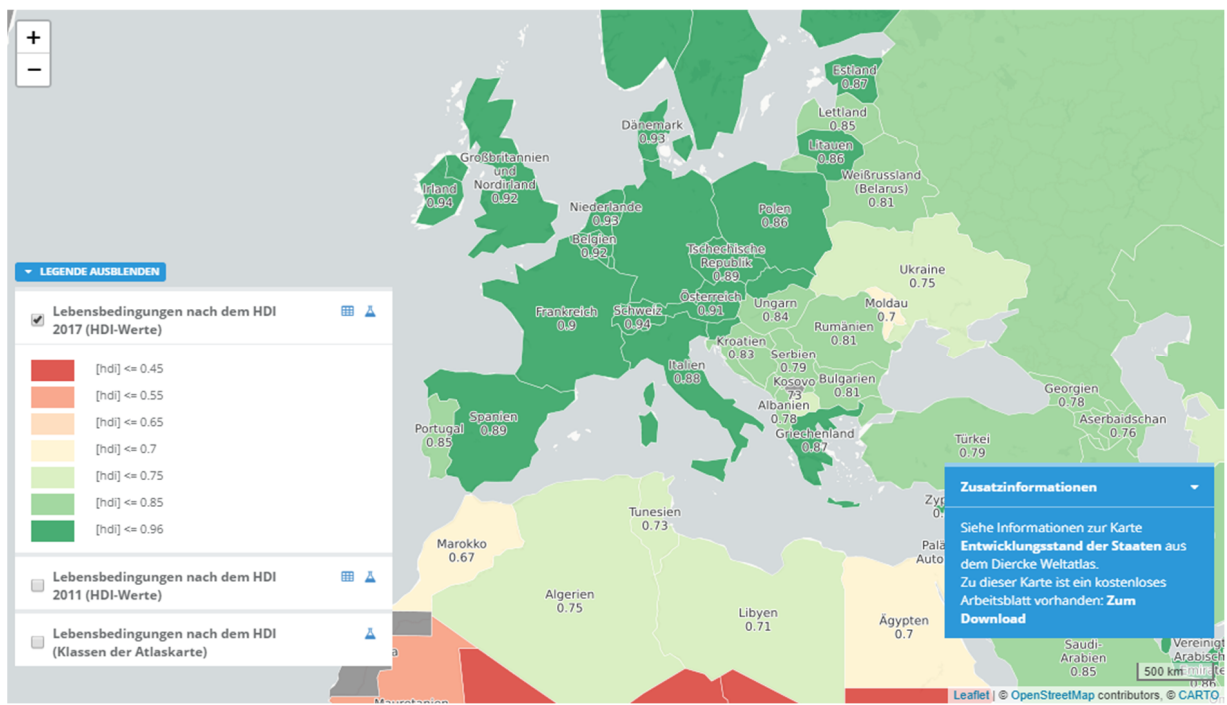

**Figure 5.** The example of the map made with Diercke GIS [41].

*The third stage* of GIS technologies development in school geographical education is related to the spread of school GIS education in Eastern Europe, Asia, Africa and Latin America during 2005–2012.

French educational system does not see GIS as a priority. In 1990, the Ministry of Education attempted to introduce a program to study GIS and remote sensing at the national level, but this program did not last long. To change the situation with GIS education, ESRI developed software in French. The Ministry of Education together with the National Institute of Geography (NIG) created the EDUGEO website on the French geoportal to open access to geographic information resources for teachers and students [14].

Presently we observe a cardinal change in GIS education in European Union. In 2014 ESRI signed an agreement with representatives of the EU education system which brought about innovative solutions in European Union education within three years, particularly, the application of geographic information systems in teaching. The ArcGIS Online platform as a base provides lesson plans and other methodological and organizational materials. The other example of this approach is the Schoolnet Future Classroom Lab project launched in 2011, and is specifically designed for online learning via which students and teachers get the ArcGIS online platform access. Such online educational services fit perfectly into modern realities and make an affordable learning service that makes study easier for both students and teachers.

Turkey, India and China take the leading position in GIS-education among Asian countries. Although GIS education in Turkey was developed much later than in USA, Canada, Great Britain, etc., today geo-information technologies are actively implemented at all educational levels. Thus, the national geography programs in Turkey, introduced by the Ministry of National Education in 2005, recommended the introduction of new technologies, including GIS technologies in geographical classes. Since then, the country has published methodological materials for secondary schools, held various conferences and educational programs for teachers of geography.

The GIS study in India was introduced into the high school curriculum in 2000, as part of the National School Curriculum in 2000. It led to the creation of appropriate curricula and publication of teacher's textbooks, the development of teacher training, and the purchase of software and technical equipment.

China widely involves GIS in higher education. In 2004, GIS was studied at 120 universities and colleges, and today this figure raised up to 150 universities. Despite China's school education reform and the introduction of a new progressive school curriculum in 2004–2005, GIS was not widely implemented in it. Meantime in Taiwan new educational program GIS education is marked as one of the priority areas after the transition to 12-year education in the period 2006–2009.

We want to pay special attention to the introduction of GIS in the educational process of African schools. The main problem of the black continent schools is extremely poor technical equipment. The lack of computers, one of the main components of GIS, is the biggest deterrents to the development of geographic information technology in African schools. Hence, it is of great interest the experience of researchers from the Faculty of Geography, Geo-informatics and Meteorology at the University of Pretoria (South Africa), who under the auspices of ESRI developed a paper-based GIS Education Package for South African schools with small information resources [10]. The teaching material complex includes a topographic map at a scale of 1 : 50000, ortho-photos at a scale of 1 : 10000, tracing paper, proofreader, colored chalk, glue, notebooks for students and a guide for teachers. The textbook's list of contents amounts seven practical lessons:

Lesson 1 Theme: Introduction into GIS.
Lesson 2 Theme: Determinations of GIS, GIS components and GIS application.
Lesson 3 Theme: Raster and vector data application in GIS.
Lesson 4 Theme: GIS data input.
Lesson 5 Theme: Dots, lines and polygons digitizing in GIS.
Lesson 6 Theme: Digitizing and Buffering.
Lesson 7 Theme: Geographic data analyses.

*The fourth stage* of GIS technologies development in school geographical education is characterized by the spread of GIS in school curriculums in most countries (from 2012 to the present).

Russia takes one of the leading places among the East European countries in GIS secondary education. Aleksandr M. Berlyant [37], Aleksei I. Krylov [30] and Vladimir S. Tikunov [17] researches laid a foundation for GIS realization at higher education in Russia. However, the lack of the necessary methodology, staff and facilities substantially restrains the primary GIS-education in the country.

Russia currently works on developments to create a school GIS, which aims to solve new and traditional geographical tasks in geography lessons. The school GIS "Living Geography" is very popular in Russian schools. It is an educational and methodical complex that contains a software shell with tools for working with geospatial data, sets of digital geographical and historical-geographical maps, a set of space images and guidelines for teachers. It enables a student under the guidance of a teacher to determine their location or geographical coordinates of objects on the ground, build a digital map of an area, draw a route on a digital map or plan, etc.

In Ukraine, there is the need to introduce students to the basics of geographic information systems. The GIS Association of Ukraine was established in 1996. Experts of the association took the first steps in the direction of training personnel with GIS-outlook and capable of spreading geospatial analysis technology. They held scientific and methodical seminars, published scientific articles and monographs. Nevertheless, for objective and subjective reasons geographic information systems did not find a worthy application in school education.

Considerable changes took place in this sphere only in recent years. This is primarily due to the profiling of the school in accordance with the Concept of General Secondary Education, approved by the Resolution of the Board of the Ministry of Education and Science of Ukraine and the Presidium of the Academy of Educational Sciences of Ukraine in 2001. This created much better conditions for differentiated learning, taking into account the individual characteristics of students' development. As a result, Ukrainian researchers have developed a number of elective course programs for specialized classes.

Liudmyla M. Datsenko and Vitalii I. Ostroukh together with the editorial board of the State Research and Production Enterprise "Kartographia" developed the program of the optional course "Fundamentals of Geographic Information Systems and Technologies". The program is designed to study the basics of geographic information systems and technologies in 10–12 grades of senior school of all profiles of natural-mathematical, technological directions. The purpose of the course "Fundamentals of Geographic Information Systems and Technologies" is to provide students with a theoretical geo-information knowledge base, skills to use effectively modern geographic information systems and technologies. It is supposed to provide them with basic principles of geographic information representation and visualization in geo-information systems and skills in modern geographic information software [13].

Liudmyla M. Datsenko published a textbook "Fundamentals of Geographic Information Systems and Technologies" in 2013 to implement practically the analyzed program. The material of the textbook helps students to get a theoretical knowledge base on the basics of geo-informatics, skills and abilities to use effectively modern geo-information systems and technologies [15].

Olha P. Kravchuk proposed one more variant of the optional course "Geographic information systems". The course extends and elaborates "Geography" and "Informatics" general education subjects, deepens senior students' ideas about GIS, and how to apply GIS in scientific research. The course content provides variety for the profile schools and encourages professional orientation. The program trains students to perform practical tasks both in the classroom and independently after classes [29].

The contribution of the staff of the Institute of Advanced Technologies under the leadership of Oleksandr V. Barladin to school GIS education must be also credited. Thus, they developed the educational IPT-UCHGIS, which has a wide range of functional capabilities to support the educational process, including solving many practical problems, geo-information support of educational practices, expeditions and excursions. We would like to mention the administrative names module of search, a drawing module for the marking-off various polygonal, linear or dotted objects, relief visualization, a profile plotting, etc. (figure. 6). It makes possible to scale, edit the properties of objects (colors, thicknesses, names), and print the created maps.

The Concept of the new Ukrainian school announces the importance of ICT and digital competence formation: "ICT and digital competencies envisage confidence and critical appraisal in the use of Information and Communication Technology (ICT) to produce, research, process and exchange information at the workplace, in the public domain and in personal communication. Information and media competence, the fundamentals of programming, algorithmic thinking, working with databases, and skills in Internet security and cyber security. Understanding of the ethics in information processing (copyright, intellectual property, etc.)" [21].

Consequently, radical changes occur in GIS-education of the secondary school. Thus, the regulations that define the strategy for the development of geographical school education state that GIS is a perspective area for the development of computer technology to be used in curricula. In geography lessons, students need to understand that geographic information systems associate a cartographic object that has a shape and location with descriptive and attributive information that belongs to those objects and characterizes their properties. As a result, the new Geography curriculum of the secondary schools (grades 6–9), the course "Ukraine in the world: nature, population" (grade 8), incorporates the study of GIS technologies of cartographic Internet sources, geo-information and modern navigation systems in geography lessons. The study of electronic maps and globes, cartographic Internet sources, navigation maps, geographic information systems, remote sensing of the Earth and areas of their practical application is included in the course "Geographical space of the Earth" (11 grade standard level and profile level). Yet, the issue of form, technical means, programs and methods to fulfill the requirements of the curriculum for a teacher remains open.

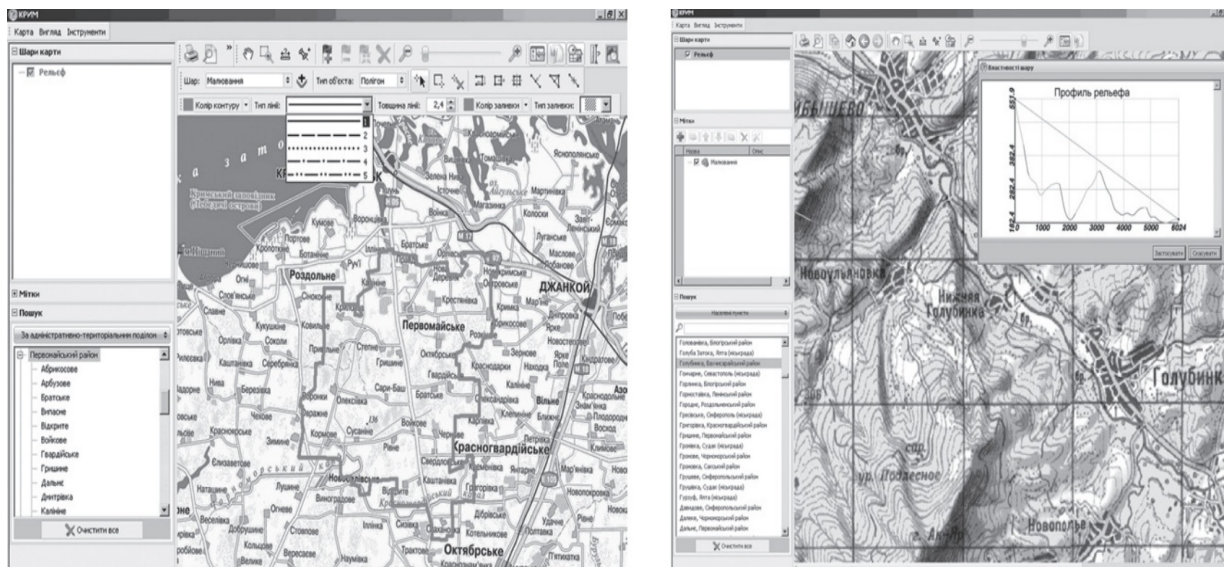

**Figure 6.** Searching by administrative units, relief visualization and making a hypsometric profile in IPT-UCHGIS school-and-student program [4].

## 3. Conclusions

A retrospective analysis of using GIS in global school practice allows to make the following conclusions.

1. The analysis of the sources on the research problem reveals the main stages of development of GIS technologies in world schools:
    - the first stage (late 1970 – early 1990's) marked the development of the technology as well as the emergence of the first GIS training programs in the leading universities of the United States and Canada, which led to partnership agreements between schools and universities the of these countries;
    - the second stage (mid-1990s – beginning of the 21$^{st}$ century) is the dissemination of GIS-educational programs in schools of the most developed countries of Europe (France, Germany, Great Britain, etc.) and Australia, as well as active participation of leading GIS developers packages (ESRI, Intergraph, MapInfo Corp., etc.). in GIS education at all levels;
    - the third stage (2005–2012) is related to the spread of the GIS school education in Eastern Europe, Asia, Africa and Latin America;
    - the fourth stage (from 2012 to the present) represents the appearance of geographic information systems in school curricula of most countries.

2. The active dissemination of GIS technologies in education demonstrates the methodological problems of GIS education: reasonable and competent use of world experience in the methodology of implementing GIS in the school educational process, but taking into account the specifics of the educational process of their country, is the key to obtaining a decent pedagogical result.

3. The lack of specialized educational GIS programs in most countries, adapted to the school program, forces schools to use professional GIS in the framework of special educational programs carried out by large GIS companies, but a modern teacher is not always able to understand these complex software packages.

4. Judging from to world experience, the successful development of school GIS education is observed only in those countries where there is a system of interactive teacher education through professional courses, meetings, conferences, congresses, etc.

5. We see further development of research in the development of educational and methodological support for the introduction of GIS in the school educational process, determining the didactic foundations of teaching the basics of GIS, taking into account the retrospective analysis.